\documentclass[twocolumn,prl,epsfig,showpacs]{revtex4}

\usepackage{graphicx}

\begin{document}

\setlength{\unitlength}{1cm}
\newcommand{\be}{\begin{eqnarray}}
\newcommand{\ee}{\end{eqnarray}}
\newcommand{\bee}{\begin{eqnarray*}}
\newcommand{\eee}{\end{eqnarray*}}
\newcommand{\I}{\mbox {\sc 1}}
\newcommand{\T}{\mbox {\sc T}}
\newcommand{\p}{\mbox {\sc P}}
\newcommand{\asy}{{\cal O}}
\newcommand{\ind}{\hskip 0.5cm}


\title {Critical conditions for a stable molecular structure}

\date{\today}

\author {Vincenzo Grecchi}
\author {Andrea Sacchetti}

\affiliation
{Dipartimento di Matematica, Universit\'a di Bologna, P.za di S.Donato 5, Bologna 40127, Italy\\
Dipartimento di Matematica, Universit\`a di Modena e Reggio Emilia, Via Campi 213/B, Modena 41100, Italy}

\email{Grecchi@dm.unibo.it, Sacchetti@unimo.it}

\begin {abstract}
Here, we show how the molecular structure appears and becomes stable for supercritical physical conditions. \ In particular we consider,  for the ammonia molecule in a gas, a simple model  based on a standard non-linear double-well Schr\"odinger equation with a dissipative term and a term representing weak collisions.
\end{abstract}


\maketitle


 \ind The existence of a well defined molecular structure for a symmetric molecule is an old and ongoing problem in chemistry \cite {W}, \cite {V}. \ Various attempts have been made to make quantum mechanics agree with apparently stationary asymmetric states. \ It was clear from the beginning that  the action of the environment would be the basic reason of this effect, since from a quantum mechanics point of view an isolated symmetric molecule has no structure. \ The point is to find the simplest model simulating the influence of the environment, and to exploit it.

\ind In 1927 Hund \cite {H} understood that in the case of certain molecules, such as ammonia $NH_3$, the basic model is a double well one, this simple model is able to give the splitting and the inversion frequency of an isolated molecule, but not the structure related to localization. \ From the experimental point of view, experimental data \cite  {BL} showing the decrease of the splitting, as a decrease of the inversion line (red shift), for increasing values of the gas pressure appeared in 1948. \ Theoretical models for inter-molecular interactions were considered, and they show an interesting fit of the data at small pressure \cite {M}, \cite {BR}.

\ind Further researchers proposed some simple models in order to explain the localization phenomenon of a symmetric molecule \cite {D}, \cite {CJ}. \ Later, some works \cite {GMS1}, \cite {GS} have exploited the strong effect of metastability, both static and dynamic, on red shift and localization.

\ind In 1995 it was shown \cite {GM} that a non-linear Schr\"odinger equation, previously  suggested by Pratt \cite {P} for the hydrogen case, is able to give spontaneous symmetry breaking and the bifurcation of the ground state at the value
\be
\mu_c =1 \label {eq1}
\ee
of the dimensionless non-linearity parameter $\mu$ defined below (which is expected to be monotically dependent on the gas pressure). \ Later, it became clear that this non-linear model is able to give the red shift and localization too. \ In particular, in \cite {V} and \cite {GMS} it is suggested that localization appears for $\mu =2\mu_c$; while in \cite {JPT} it is suggested that the critical value (\ref {eq1}) found in \cite {GM} coincides with the vanishing of the inversion line,  and  corresponds to the pressure of about 1,7 atmospheres at room temperature \cite {nota1}.

\begin{figure}
\includegraphics[height=8cm,width=8cm]{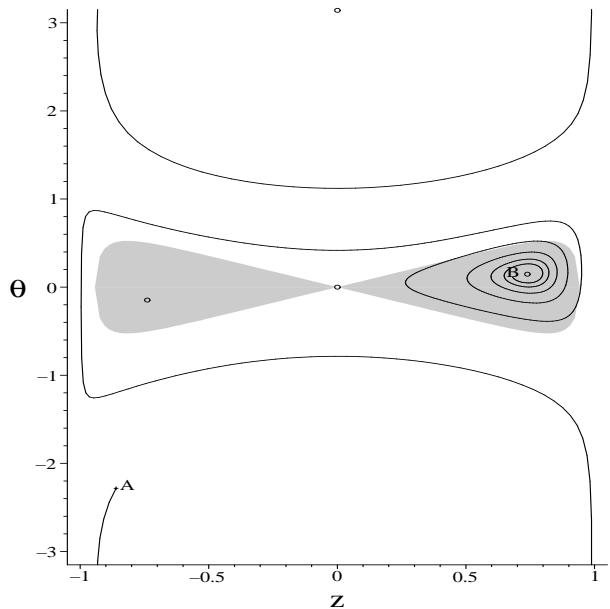}%
\caption{\label {fig1} In this figure we plot the parametric graph of the solution $z(t),\ \theta (t)$ of equation (5) for a given
initial condition (point A) in the case $\mu = 1.5$, $\zeta = 0.2$, point $B$ denotes the position of the state at $t=10 \tau$. \ Circle points denote the stationary solutions. \ In the non-dissipative case ($\zeta =0$) we have only {\it vibrational} motions around one well inside the grey region, outside the grey region; we have periodic {\it beating} motions between the two wells. \ Here, $z$ denotes the imbalance function taking values in the interval $[-1,+1]$ and $\theta$ is a torus variable taking value in the torus $(-\pi ,+\pi]$.}
\end{figure}

\begin{figure}
\includegraphics[height=8cm,width=8cm]{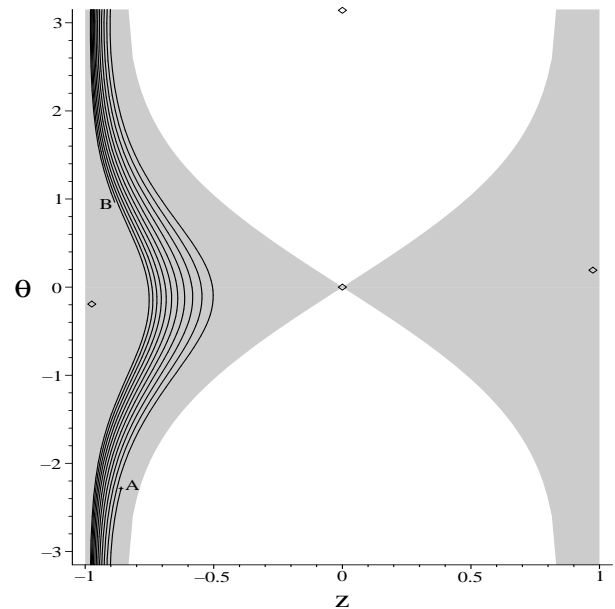}%
\caption{\label {fig2} In this figure we plot the parametric graph of the solution for a given
initial condition (cross point) in the case $\mu = 4.5$, $\zeta =0.2$. \ Diamond points denote the stationary solutions.}
\end{figure}

\ind In this letter we consider the non-linear double-well model as above \cite {GM} in order to obtain the stability of the localization for finite values of the parameters. \ Let us recall \cite {Wi} that actually we consider a substituted ammonia molecule $NHDT$, so that a localized state has one of the two possible chiral configurations. \ In the approximation of a two-level system, an invariant quantity ({\it energy}) appears, and we observe periodic motions of two kinds: {\it vibrational} periodic motions around just one well (inside the grey regions of Fig. \ref {fig1} and \ref {fig2}) or {\it beating} periodic motions between the two wells. \ In order to have a more physical model, we add a dissipative term, related to the photonic emission in the radio frequency range, giving the decreasing of the {\it energy}. \ We also take into account the effect of weak collisions. \ As a result we find that a chiral configuration is unstable if the non-linearity parameter is less than $3\mu_c$. \ For larger values of this parameter, the chirality states become stable provided that the interval between two collisions is larger than the relaxation time, in order that when a collision occurs, the state be near an asymmetric stationary state.

\ind In order to better understand the situation, consider Figs. \ref {fig1} and \ref {fig2} where the space of states is shown, as a sphere projected on a square by a Mercator map. \ The two coordinates $z$ and $\theta$ respectively represent the imbalance variable (which measures the  localization) and the relative phase of the components of the state with respect to the two localized states. \ A collision actually changes $\theta$ but not $z$. \ Thus, for $\mu$ greater than $3\mu_c$ (see Fig. \ref {fig2}), the line defined by a value of $z$ equal to the one of  an asymmetric stationary state is fully contained in the {\it vibrational} region. \ In contrast, for $\mu < 3 \mu_c$ (see Fig. \ref {fig1}) we have that a change of $\theta$, due to a collision, could shift an asymmetrical stationary state into the {\it beating} region.

\ind The Hamiltonian for a single ammonia molecule takes the form $ H_0 = - \frac {\hbar^2}{2 m} \Delta + V $ where $V$ is a double-well potential invariant under a coordinate reflection
\bee
V = \p V \p ,
\eee
where $\p$ is the unitary symmetry operator representing the inversion of the $n$--th coordinate
\bee
(x' , x_n)  \to ( x' , -x_n) , \ x' = (x_1 , \ldots , x_{n-1}), \ n \ge 1 .
\eee
The tunneling time through the inter-well potential barrier is inversely proportional to the energy splitting $\Delta E$
between the odd-- and even--parity eigenstates $|-\rangle $ and $|+\rangle $ with energies  $\lambda_-$ and $\lambda_+$, and the solution of the unperturbed equation
\bee
i \hbar \frac {\partial }{\partial t} | \psi (t) \rangle  = H_0 |\psi (t) \rangle
\eee
shows a beating motion between the two wells with period
\bee
\tau = \frac {2\pi \hbar}{\Delta E },\ \ \Delta E =  \lambda_- - \lambda_+ .
\eee
The actual semi-classical parameter is the energy splitting $\Delta E$, and we choose the units such that $\hbar =1$ and $\Delta E \ll 1$.

\ind Let us consider the interaction of a single molecule with the other molecules of the gas. \ In the mean field approximation, the new Hamiltonian takes the form
\bee
H = H_0 + W
\eee
where the term $W$ is given by the polarization of the external environment due to the presence of the single ammonia molecule itself. \ Because of the dissipative terms, the perturbation $W$ could be written by means of a non-linear and non-Hermitian term
\bee
W  = \nu g \left ( \epsilon \I + i \eta \p \right ), \ \ \nu = \langle \psi (t) | g | \psi (t) \rangle , \ \ \epsilon , \ \eta <0,
\eee
where $g(x)$ is a given odd function, $\p g \p =-g$, and where the parameters $\epsilon$ and $\eta$ respectively measure the strength of the dipole-dipole and the dissipation interactions. \  As in the case of   complex Ginzburg-Landau equations \cite {AK} and of Gross-Pitaevskii equations with weakly dissipative effects \cite {CV}, we expect to observe vortex solutions here.

\ind We underline that in the non-dissipative case, where $\eta =0$, then we have the conservation of the charge:
\be
\langle \psi (t)|\psi (t) \rangle  = \| \psi (t) \|^2 =1
\label {eq2}
\ee
and the conservation of the {\it energy} functional
\bee
{\cal E} (\psi ) = \| \nabla \psi \|^2 + \langle \psi |V|\psi  \rangle + \frac {1}{2}\epsilon \langle \psi |g|\psi  \rangle^2.
\eee
In the dissipative case, where $\eta <0$, since the term $W$ is such that $[\T \p ,W]=0$, where $\T |\psi (t) \rangle  = | \psi (t) \rangle^\star $ is the time-reversed operator, we still have the conservation of the charge (\ref {eq2}), and we expect that the {\it energy} functional decreases. \ The system then relaxes towards local minima of the functional.

\ind Here, we make use of the two-level model where we approximate the wave-function $|\psi (t) \rangle $ by means of its projection on the two lowest states. \ In particular, in such an approximation the total wavefunction of the system may be expanded as
\be
|\psi (t)\rangle  = a_R (t) |R \rangle  + a_L (t) |L \rangle , \label {eq3}
\ee
where
\bee
|R \rangle = \frac {1}{\sqrt 2} \left ( |+\rangle + |-\rangle \right ) , \
|L \rangle = \frac {1}{\sqrt 2} \left ( |+\rangle - |-\rangle \right )
\eee
are the right and left hand-side states; they are such that $\p |R\rangle  = |L \rangle $. \ The normalization condition (\ref {eq2}) on the wavefunction $|\psi (t) \rangle$ implies that $|a_R|^2+|a_L|^2=1$.

\ind By substituting $|\psi (t) \rangle $ by (\ref {eq3}) in the time-dependent Schr\"odinger equation
\bee
i \hbar \frac {\partial }{\partial t} | \psi (t) \rangle  = H |\psi (t) \rangle
\eee
it follows that the expansion coefficients $a_R$ and $a_L$ have to satisfy to the following system of ordinary differential equations
\be
\left \{
\begin {array}{lcl}
i \dot a_R &=& \Omega  a_R - \omega a_L + \epsilon \nu c a_R +i \eta \nu c a_L \\
i \dot a_L &=& \Omega  a_L - \omega a_R - \epsilon \nu c a_L -i \eta \nu c a_R \\
\nu &=& c (|a_R|^2 - |a_L|^2 )
\end {array}
\right.  \label {eq3a}
\ee
where
\bee
\omega = \frac 12 (\lambda_- - \lambda_+ ) ,\ \ \Omega = \frac 12 (\lambda_- + \lambda_+ )
\eee
and where we set
\bee
c= \langle R | g | R \rangle =  -\langle L | g | L \rangle .
\eee

\ind In order to study the beating motion, it is convenient to introduce the relative phase
\bee
\theta = \arg (a_R) - \arg (a_L)  ,
\eee
which is a torus variable, and the imbalance variable
\bee
z = |a_R|^2-|a_L|^2,
\eee
which takes values in the interval $[-1,+1]$. \ They have to satisfy to the system of ordinary differential equation
\be
\left \{
\begin {array}{lcl}
\dot z &=& \omega Z (z,\theta )  \\
\dot \theta &=& \omega \Theta (z, \theta )
\end {array}
\right. \label {eq4}
\ee
where
\bee
Z (z, \theta ) =  2 \sqrt {1-z^2} \left [ \sin \theta  - \zeta z \cos \theta \right ]
\eee
and
\bee
\Theta (z, \theta ) = - 2\frac {z}{\sqrt {1-z^2}} \left [ \cos \theta + \zeta z \sin \theta \right ] + 2 \mu z \eee
and
\bee
\mu = -c^2 \epsilon /\omega  , \ \ \zeta = -c^2 \eta /\omega .
\eee
Here, $\mu$ and $\zeta$ represent positive dimensionless parameters that measure the effective non-linearity and the dissipation, respectively.

\ind We observe the existence of the critical value (\ref {eq1}) for the non-linearity parameter $\mu$. \ We consider, at first, the {\it weak non-linearity} case such that $ \mu < \mu_c$. \ In such a case, equation (\ref {eq4}) admits just two stationary solutions corresponding to the unperturbed even-- and odd--parity  eigenstates: $z_1=0$ and $\theta_1 =0$, corresponding to the even--parity eigenstate, is a {\it stable} stationary solution, and $z_2=0$ and $\theta_2 =\pi $, corresponding to the odd-parity eigenstate, is an {\it unstable} stationary solution.

\ind Then, we consider the {\it strong non-linearity} case such that $\mu >\mu_c$. \ We observe that when $\mu $ takes the value $\mu_c$ then the stable stationary solution makes experience of a bifurcation phenomenon \cite {GM}. \ More precisely, for $\mu >\mu_c $ we have 4 stationary solution; two of them still correspond to the unperturbed even-- and odd-parity eigenstate, the other two correspond to asymmetrical states that, in the limit of large non-linearity, are fully localized on just one of the two wells. \ The stationary solution $z_1=0$ and $\theta_1 =0$, corresponding to the even--parity eigenstate, is a {\it saddle point} for $\mu >\mu_c$;
the stationary solution $z_2=0$ and $\theta_2 =\pi $, corresponding to the odd--parity eigenstate, is still an {\it unstable} solution; $z_3=\sqrt { (\mu^2 -1)/(\mu^2 + \zeta^2)} $ and $\theta_3 = \mbox {arctan} (\zeta z_3)$, and $z_4=-z_3$ and $\theta_4 =-\theta_3$ are {\it stable} asymmetrical stationary solutions. \ Thus any state generically goes near to one of these two asymmetrical stationary states.

\ind Therefore, in the strong non-linearity case we have that any initial state, except the two even-- and odd-parity unperturbed eigenstates, finally goes to one of the two asymmetric stationary eigenstates giving a chiral configuration for the ammonia molecule.

\ind Now, we show that this chiral configuration is stable with respect to collisions when the pressure is large enough. \ We underline that in the ammonia case the thermal energy at room temperature is smaller than the distance between the doublet $\{ \lambda_\pm \}$ and the other energy levels, so that the validity of the two-level approximation holds, and it is much larger than the splitting energy, so that a collision could produce a strong variation of the {\it energy} ${\cal E} (\psi )$.

\ind To this end we introduce a simplified model for molecular collision. \ When the single molecule undergoes a collision we add to the Hamiltonian $H$ a perturbative term of the type $ f (x) v (t)$ where $f(x)$ is a function with compact support and $v(t)$ is a given time dependent function. \ For instance, let $v(t ) = \chi_{t_1 , t_2} (t)$ where $\chi $ is the characteristic function on the interval $[t_1,t_2]$, and where we assume that the perturbation acts for a time much shorter than the beating period, that is
\be
t_2-t_1  \ll \tau, \label {eq5}
\ee
Equation (\ref {eq3a}) then takes the form
\bee
\left \{
\begin {array}{lcl}
i \dot a_R &=& \Omega  a_R - \omega a_L + \epsilon \nu c a_R +i \eta \nu c a_L + v(t) c_R a_R \\
i \dot a_L &=& \Omega  a_L - \omega a_R - \epsilon \nu c a_L -i \eta \nu c a_R + v(t) c_L a_L \\
&& \ \ c_R = \langle R | f | R \rangle , \ c_L = \langle L | f | L \rangle
\end {array}
\right.
\eee
since
\bee
\langle R | f | L \rangle = \overline {\langle L | f | R \rangle} \sim 0
\eee
Hence, the system (\ref {eq4}) takes the form
\bee
\left \{
\begin {array}{lcl}
\dot z &=& \omega Z (z,\theta )  \\
\dot \theta  &=& \omega \Theta (z, \theta )
+ (c_R-c_L) v (t )
\end {array}
\right.
\eee
from which it follows that $|z(t_2 )-z(t_1 )| \ll 1$, since (\ref {eq5}), and where it is not possible obtain a similar bound for $\theta$ when $c_R \not= c_L$.

\ind This fact, that is the relative phase is strongly modified after a generic collision (such that $c_R\not= c_L$), does not actually destroy the chiral configuration of a localized ammonia molecule if the pressure is large enough, i.e. such that $\mu >3\mu_c$. \ Indeed, in such a case we have that the stable solution $(z_3,\theta_3 )$ (respectively $(z_4,\theta_4 )$) has a basin of attraction containing the strip $z \ge z^\star $ (respectively $z \le - z^\star $) if $\zeta$ is small enough and where $z^\star = 2\sqrt {2}/3<z_3$. \ We explain this fact by means of a continuity argument in the limit of $\zeta =0$. \ Indeed, in such a limit we have the existence of two separatrix lines  \cite {GMS}, \cite {AV} starting from the stationary solution $(z_1, \theta_1 )$ and satisfying the equation
\bee
\sqrt {1-z^2}\cos \theta + \frac 12 \mu z^2 =1.
\eee
It is not hard to see (Fig. \ref {fig2}) that these paths are contained in the strip
 $-z^\star \le z \le z^\star $ if $\mu >3\mu_c $. \ As a result, it follows that a perturbation due
 to a collision acting in an interval of the order (\ref {eq5})  shifts a state initially near to
  one of the asymmetric stationary stable eigenstates, to a state belonging to the basin of
  attraction of the stationary eigenstate itself. \ In particular, the state will be always
  far enough from the unstable stationary state and thus, after a finite time depending on $\zeta$,
  it returns near to the initial stationary state without visiting the other well, provided that in this
  period another collision does not occur.

\ind Finally, let us notice that the molecular structure is not completely stable for strong collisions;
indeed we have the phenomenon of racemization which makes the statistical mean of the chirality vanish at
large time, and that an external electromagnetic field in the radio frequency range could be able to destroy
the molecular structure.

\ind In conclusion, in this letter we have shown that a dissipative non-linear model is able
to explain the molecular structure of symmetric molecules. \ Such chiral configurations are stable
for weak collisions provided that the non-linearity parameter is larger than the critical value $3\mu_c$,
 and that the frequency collisions is small enough. \ The kind of trap we propose here is simple, but not
  trivial, and in any case is able to make stable the spontaneous symmetry breaking given by the non-linearity.
  \ It could be relevant in the theory of dechoerence, related to the appearance of classical mechanics,
  and in the study of many irreversible precesses. \ As it clearly appears in this letter, we have only considered
  the spontaneous symmetry breaking effect. \ In fact, in the case of organic molecules, one enantiomer may be dominant. \
  This effect could be
  explained by means of a small initial enantiomeric excess, due to the parity violation for weak interactions \cite {Q},
  largely amplified during a very long time by the dynamics of the system \cite {K}.

\begin {acknowledgments}
This work is partially supported by the Italian MURST, INFN and INDAM-GNFM.  We would thank Prof. Andr\'e Martinez
for helpful discussions and remarks.
\end {acknowledgments}

\end {document}